**Jannick P. Rolland**
Optical Diagnostics and
Applications (ODA) Lab
University of Central Florida
College of Optics and Photonics,
Institute of Simulation and Training,
and School of Computer Science
University of Central Florida
4000 Central Florida Boulevard
Orlando, FL 32816
Rolland@odalab.ucf.edu

**Frank Biocca**
Media Interface and Network
Design (MIND) Labs
Michigan State University
East Lansing, MI 48824

**Felix Hamza-Lup**
Optical Diagnostics and
Applications (ODA) Lab
School of Computer Science
University of Central Florida
4000 Central Florida Boulevard
Orlando, FL 32816

**Yanggang Ha**
**Ricardo Martins**
Optical Diagnostics and
Applications (ODA) Lab
College of Optics and Photonics
University of Central Florida
4000 Central Florida Boulevard
Orlando, FL 32816

# Development of Head-Mounted Projection Displays for Distributed, Collaborative, Augmented Reality Applications

## Abstract

Distributed systems technologies supporting 3D visualization and social collaboration will be increasing in frequency and type over time. An emerging type of head-mounted display referred to as the head-mounted projection display (HMPD) was recently developed that only requires ultralight optics (i.e., less than 8 g per eye) that enables immersive multiuser, mobile augmented reality 3D visualization, as well as remote 3D collaborations. In this paper a review of the development of lightweight HMPD technology is provided, together with insight into what makes this technology timely and so unique. Two novel emerging HMPD-based technologies are then described: a teleportal HMPD (T-HMPD) enabling face-to-face communication and visualization of shared 3D virtual objects, and a mobile HMPD (M-HMPD) designed for outdoor wearable visualization and communication. Finally, the use of HMPD in medical visualization and training, as well as in infospaces, two applications developed in the ODA and MIND labs respectively, are discussed.

## I    Introduction

With the increase in multidisciplinary workforce and the globalization of information exchange, multiuser work teams are often distributed. Organizations increasingly seek ways to effectively support these teams by allowing them to share and mutually interact with 3D data in a common distributed workspace. As a result there has been increased interest in and reliance upon the use of distributed systems technologies that support multiuser distributed 3D visualization. Furthermore, economic, political, and health concerns may also fuel increased reliance on distributed work environments.

The design of mobile and distributed augmented reality (AR) systems is best driven by concrete real world applications testable in real world environments. In this paper we review a research and development program to create and test AR displays and interface designs that support local, distributed, and mobile teams that (1) require immersive interaction with large scale 3D visualizations, but also (2) have full sensory awareness of the physical environment around them, and (3) are able to see and interact face-to-face with local and remote participants. Example applications include collaborative and distributed science and engineering design via 3D visualization, 3D medical visualization and training, effective face-to-face distributed business negotiation and conferencing, distance education, and mediated governmental services. The displays and interfaces are designed to support mobile AR navigation, telepresence, and advanced information systems, which we refer to as mobile infospaces, for applications such as medical and disaster relief scenarios.

The AR systems introduced below aim to create a compelling sense of fully embodied interaction with spaces and people that are not immediately present in the same physical environment. The acceptance and efficiency of distributed collaborative systems may require advanced media technologies that can create a strong sense of social presence, defined as the sense of "being with others" (Short, Williams, & Christie, 1976; Biocca, Harms, & Burgoon, 2003). Participants may need to feel that people who are not in the same room or place have a similar impact on team processes as those who are physically present. To create this effect, our research efforts include the development of technologies that enable both 3D dataset visualization (Hamza-Lup, Davis, & Rolland, 2002), together with 3D face-to-face interaction among distributed users (Biocca & Rolland, 2004), and algorithms to synchronize shared states across distributed 3D collaborative environments (Hamza-Lup & Rolland, 2004a, 2004b). To verify the effect, we have developed ways to measure the degree to which teammates feel socially present with remote collaborators. Because there are many acronyms in this paper, we provide a list in Table 1 to assist the reader.

In this paper, we first discuss the relative advantages of head-mounted projection displays (HMPDs) for distributed 3D collaborative AR systems. We then review differences between eyepiece and projection optics, and discuss various trade-offs associated with the technology. The exploration of various HMPD designs within our laboratories is presented as well as two novel emerging technologies based on the HMPD, the teleportal HMPD (T-HMPD), and the mobile HMPD (M-HMPD). Finally, a review of two collaborative applications, medical visualization and training, and the design of infospaces, is provided.

**Table 1.** *List of Acronyms Employed in this Paper*

| Acronym | Full phrase |
| --- | --- |
| 3D | Three-dimensional |
| ACLS | Advanced cardiac life support |
| AR | Augmented reality |
| ARC | Artificial reality centers |
| DOE | Diffractive optical element |
| ETI | Endotracheal intubation |
| FOV | Field of view |
| HMD | Head-mounted display |
| HMPD | Head-mounted projection display |
| HPS | Human patient simulator |
| LCOS | Liquid crystal on silicon |
| M-HMPD | Mobile head-mounted projection display |
| OLED | Organic light emitting display |
| T-HMPD | Teleportal head-mounted projection display |
| UIH | Ultimate intubation head |

## 2 Relative Advantages of Head-Mounted Projection Displays for Collaborative Interactions

Few technologies are well designed to support distributed work team interactions with complex 3D models. This is one key motivator for the creation of advanced image capture and display systems. 3D visualization devices, which have succeeded in penetrating real world markets, have evolved into three formats: standard monitors/shutter glasses, head-mounted displays (HMDs) (Sutherland, 1968), and projection-based displays. Projection-based displays include video-based virtual environments that use back-projection techniques to place users within the environment (Krueger, 1977, 1985) and rear projection cubes known as the CAVE (Cruz-Neira, Sandin, & DeFanti, 1993).

Each of these three common approaches currently imposes a significant increase in cost and/or limits the quality of social interaction as well as the sense of social presence of distributed team members. Monitors with shutter glasses are limited in teamwork capabilities because of the size of the display space and the lack of immersion. Among HMDs, AR displays that rely on optical see-through capability (Rolland & Fuchs, 2001) are best suited for collaborative interfaces because they enable the possibility of undistorted 3D visualization of 3D models, while they also preserve the natural interactions multiple users may have within one collaborative site. They are weak, however, at providing natural interactions among multiple users located at remote sites given that they do not support face-to-face interactions among distributed users. Also, unless ultralight weight (i.e., <10 g) optics can be designed with a reasonable FOV (i.e., at least 40°), HMDs will limit usability, regardless of their type. Technologies such as the CAVE and Powerwalls have been especially conceived for the development of collaborative environments (http://pinot.lcse.umn.edu/research/powerwall/powerwall.html). The strength of these technologies lies in the vivid sense of immersion they provide for multiple users. However, these technologies are also weak at supporting social interaction for local teams working with the models because only one user can view the 3D models accurately without perceptual distortions, leaving other members of the work team viewing distorted models. Furthermore, the technology does not support face-to-face interactions among distributed users. CAVE technology is also typically costly because of the need for multiple large projectors, and they require a large footprint, because of the rear projection on large screens.

An HMPD may be conceptually thought of as a pair of miniature projectors, one for each eye, mounted on the head, coupled with retro-reflective material strategically placed in the environment. Retro-reflective material has the property that ideally any ray of light hitting the material is redirected along its incident direction, as opposed to being diffused as common to conventional projection systems. Naturally, such an imaging scheme raises two main questions: (1) How can a projector be miniaturized to the extent that it can be mounted on the head? and (2) How can retro-reflective material provide imaging capability?

Insight into these questions may be reached by comparing the HMPD imaging capability to existing technology. Consider first conventional projector systems, such as those typically found in conference rooms. The light projected on the screen reflects and diffuses in all directions allowing multiple viewers in the room to collect a small portion of the light diffused back towards them. This small amount of diffused light enables all to see the projected images. Such projection systems require extremely high power illumination, and thus their size. They may require dimmed light in the room so that the diffusing screen is not washed out by ambient light.

Now consider how HMPD uses light. The light from the head-worn projectors is projected towards optical retro-reflective screens. The light is then redirected in a small cone back towards the source after reaching the optical retro-reflective material. This maximizes the power received by the user whose eye (i.e., the right eye for the right-eye projected image) is located within the path of the small cone of reflected light. (We shall explain the technology in further detail in Section 3.) Therefore, in spite of the low power projection system mounted on the head, bright images can be perceived by not only one user, but each user of the screen as they will not be sharing the same reflected light. Outside the path of reflection, no light will be received because the cone of light along the path is small enough to be imperceptible by the other eye of the user. Therefore no light leakage or cross talk is possible either between the eyes of one user or between users. Because there is no cross talk, such technology, interestingly, can also allow for private or secure information to be viewed in a public work setting.

The HMPD designs have evolved significantly since they were conceived. Fisher (1996) pioneered the concept of combining a projector mounted on the head with retro-reflective material strategically placed in the environment. The system proposed was biocular with one microdisplay architecture and the same image was presented to both eyes. Fergason (1997) extended the conceptual design to binocular displays. He also made a conceptual improvement to the technology consisting of adding retro-reflective material located at 90 degrees from the material placed strategically in the environment to increase light throughput. It is important to note that a key benefit of the HMPD in the original Fisher configuration is providing natural occlusion of virtual objects by real objects interposed between the HMPD and the retro-reflective material—such as a hand reaching out to grab a virtual object as further described in Section 3. Fergason's dual retro-reflective material constitutes an improvement in system illumination. However it also compromises the natural unidirectional occlusion inherently present in the Fisher HMPD.

Early demonstrations of the HMPD technology based on commercially available optics were first demonstrated by Kijima and Ojika (1997), and independently by Parsons and Rolland (1998), and (Rolland, Parsons, Poizat, & Hancock, 1998). Shortly after, Tachi and his group first developed a configuration named MEDIA X'tal which was not head-mounted, yet used two projectors positioned similarly to that in a HMPD together with retro-reflective material (Kawakami et al., 1999). He then also applied the concept to a HMPD (Inami et al., 2000).

The Holy Grail of HMD technology development is lightweight ergonomic design, and a fundamental question is whether projection optics inherently provides an advantage. Early in the development of the HMPD technology, we envisioned a mobile, lightweight, multi-user system, and, therefore, sought to develop lightweight compact optics. A custom-designed 23 g per eye compact projection optics using commercially off-the-shelf optical lenses was first designed by Poizat and Rolland (1998). The design was supported by an analysis of retro-reflective materials for imaging (Girardot & Rolland, 1999), and a study of the engineering of HMPD (Hua, Girardot, Gao, & Rolland, 2000). An ultralightweight (i.e., less than 8 g per eye) 52° diagonal FOV and color corrected projection optics was next designed using a 1.35 in. diagonal microdisplay and aspherical and diffractive optical elements (DOEs) (Hua & Rolland, 2004). The design of a 70° FOV followed using the same microdisplay (Ha & Rolland, 2004). Next a 42° FOV projection optics was designed using a 0.6 in. diagonal organic light emitting display (OLED) (Martins, Ha, & Rolland, 2003).

Such design approaches based on DOEs not only provide ultralightweight solutions, but also importantly provide an elegant design approach in terms of chromatic aberration correction. One main insight into such a correction comes from realizing that a DOE may be thought of as an optical grating that follows the laws of optical diffraction. Consequently, a DOE has a strong negative chromatic dispersion that can be used to achromatize refractive lenses. A trade-off associated with using DOEs is a potential small loss in efficiency which can affect the contrast of high-contrast targets. Thus, in designing such systems, careful quantification of DOE efficiency across the visible spectrum must be performed to ensure good image contrast and optimal performance for the system specification.

The steady progress toward engineering lightweight HMPDs is shown in Figure 1. Early in the design we considered both top- and side-mounted configurations. We originally chose a top-mounted configuration as shown in Figure 1a for two reasons: (1) To provide the best possible alignment of the optics; (2) To more easily mount the bulky electronics. Upon a successful first prototype, in collaboration with NVIS Inc. we had the main electronics board associated with the microdisplay placed away from it, in a control box, and we investigated a side-mounted display driven by a medical training application further described in Section 5.1. In this application, the trainees must bend over a human patient simulator (HPS) and moving the packaging weight to the side was desired. This approach led to the HMPD shown in Figure 1b. We learned that such a configuration successfully moved weight to the side, however it also provided some tunnel vision, as the user's peripheral vision was rather occluded by the side-mounted optics. A recent design shown in Figure 1c capitalized on ultracompact electronics associated with OLED microdisplays (e.g., http://www.emagin. com). To avoid the tunnel vision aspect which we found very disturbing, we targeted a top-mounted geometry

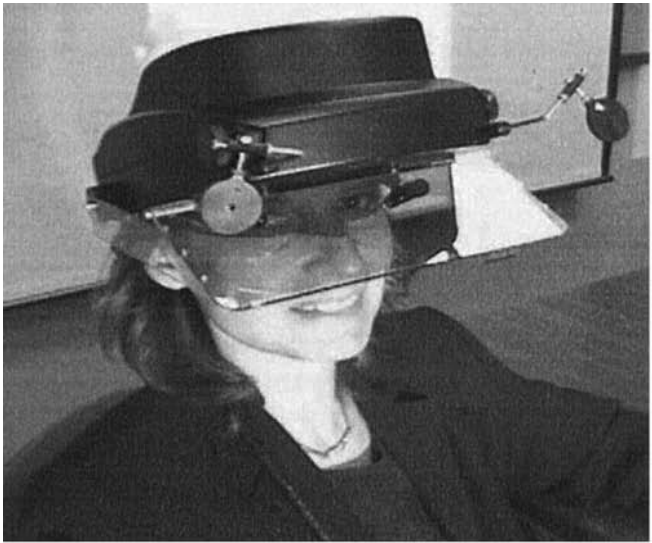

**(a)**

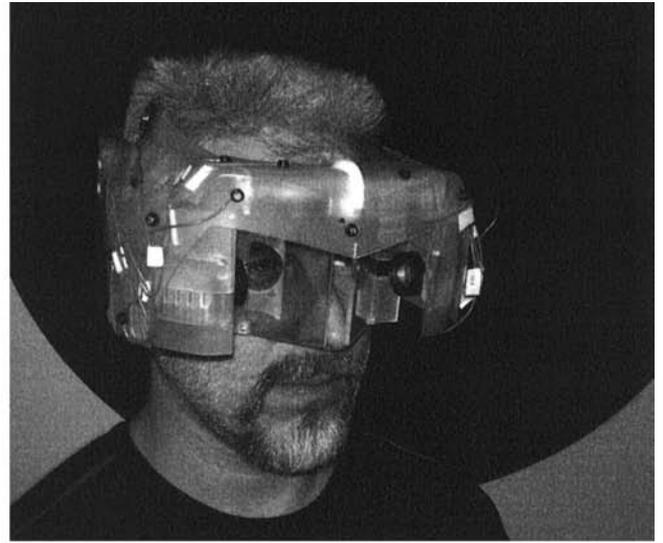

**(b)**

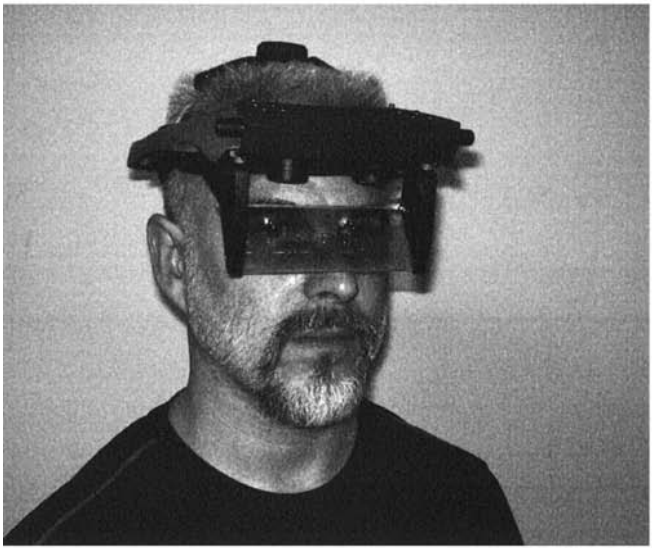

**(c)**

**Figure 1.** *Steady progress toward engineering lightweight HMPDs. (a) 2001 HMPD&FaceCapture Optics mount: Top-mounted microdisplay: AM-LCD. Pixel resolution: (640 × 3) × 480 (b) 2003 HMPD Optics mount: Side-mounted microdisplay: AM-LCD. Pixel resolution: (640 × 3) × 480 (c) 2004 HMPD Optics mount: Top-mounted microdisplay: OLED Pixel resolution: (800 × 3) × 600.*

to allow full peripheral vision. Also because of the extremely compact electronics of the microdisplay in this case, the overall system is less than 600 g, including cables, and can further be lightened past this early prototype level as metallic structures still remain. Having in-troduced the evolution of these systems, we consider the optics of these systems in greater detail in the following section because the optical design is critical to their future development making HMPDs light, bright, wide FOV, wearable, and mobile.

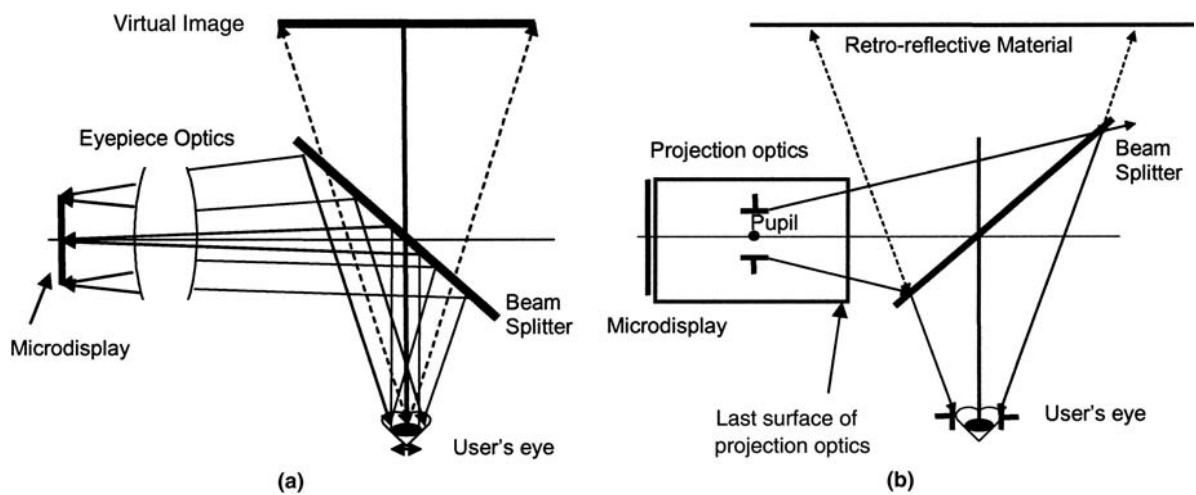

**Figure 2.** *(a) Eyepiece optics HMD: light emitted from the microdisplay reaches directly the eyes via the orientation of the beam splitter. (b) Projection optics HMD referred to as HMPD: light is first sent to the environment before being retro-reflected toward the eyes of the user, a consequence of the orientation of the beam splitter and the positioning of retro-reflective material in the environment.*

## 3. Optics of the Head-Mounted Projection Display (HMPD): Making the Displays Light, Bright, Wide FOV, Wearable, and Mobile

The optics of the HMPD consists essentially of (1) two microdisplays, one associated with each eye, and (2) associated projection optics that guides the projected images toward retro-reflective material. The optical property of such material is to retro-reflect light toward its direction of origin, or, equivalently, the eyes of the user in our optical configuration, given that the position of the eyes of the user are made conjugate to the position of the exit pupils of the optics via a beam splitter as shown in Figure 2b. Therefore, there are two unique optical components: (1) projection optics rather than eyepiece optics as used in conventional HMDs, and (2) a retro-reflective material strategically placed in the environment rather than a diffusing screen as typically used in projection-based systems, distinguish the HMPD technology from conventional HMDs and stereoscopic projection displays such as the CAVE.

## 3.1 Projection Optics vs Eyepiece Optics: Lighter, Greater Field-of-View, and Less Distortion

A comparison of HMDs based on eyepiece optics versus the projection optics of a HMPD is shown in Figure 2a,b. An important feature of eyepiece optics in a HMD is the propagation of the light solely within the HMD, between the microdisplay and the eye. In the case of projection optics in an HMPD, the light actually propagates in the real environment up to the retro-reflective material before returning to the user's eyes. The implication of this propagation scheme makes for a fundamental difference in the user experience of 3D objects and the environment. With the HMPD, it is possible to have a user occlude virtual objects with real objects interposed between the HMPD and the material, such as the hand of a user reaching out to a virtual object in an artificial reality center (ARC) display as shown in Figure 3.

The usage of projection optics allows for larger FOV (i.e., >40° diagonal) and less optical distortion (<2.5%

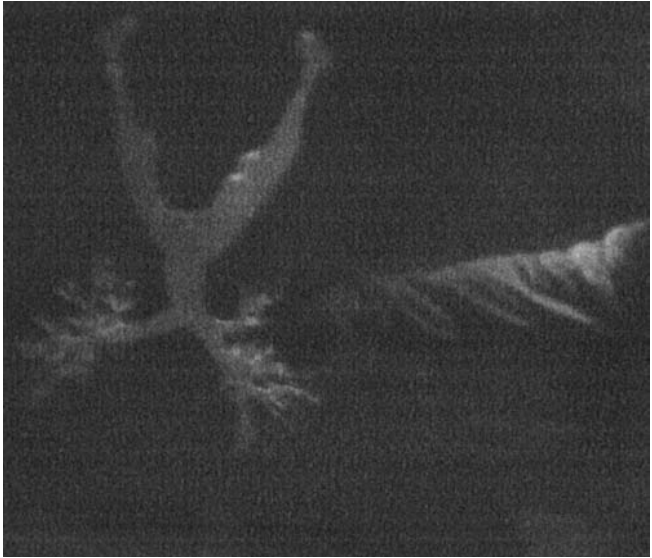

**Figure 3.** *Demonstration of occlusion capability of virtual objects by real objects: User reaches out to the trachea in the ARC display, and occludes part of the object with his hand. A grayscale image is shown here, however the display is full color. (Mandible and Trachea Visible Human Dataset 3D Models, Courtesy of Celina Imielinska, Columbia University).*

at the corner of the FOV) than obtained with conventional eyepiece-based optical see-through HMDs, for an equivalent weight. At the foundation of this capability is the location of the pupil both within the projection optics and after the beam splitter.

In conventional HMDs using eyepiece optics, one may distinguish between pupil forming and non-pupil forming systems. By pupil forming we mean that a diaphragm (i.e., an aperture stop more generally called a pupil that limits the light passing through the system) is located within the optical system and is being reimaged at the eye location. The eye location is referred to as the eyepoint because in computer graphics, a pinhole model is used to generate the stereoscopic image pairs. In the case of a pupil forming eyepiece, an aperture stop is thus located within the optics, however, the final image of this stop after the eyepiece and beam splitter is real and located outside the optics, at the chosen eyepoint for the generation of the stereoscopic images (Rolland, Ha,

& Fidopiastis, 2004). The main drawback of an external exit pupil is that as the FOV increases, the optics increases in diameter and thus the weight increases as the cube of the diameter. For the non-pupil forming eyepiece, the pupil (i.e., the image of the eye iris through the cornea) of the eye itself constitutes the limiting aperture stop of the eyepiece. However the same limitation occurs regarding the trade-off in FOV versus weight. An advantage of non-pupil forming optics, however, is that the user may benefit from a larger eye box where eye movements may more freely occur without vignetting of the image (i.e., vignetting refers to partial light loss or full loss of the image).

Projection optics is intrinsically a pupil forming optical system. In the case of projection optics, an aperture stop is located within the projection lens by design and the image of this aperture stop known as the exit pupil is virtual (i.e., it is located to the left of the last surface of the projection optics shown schematically in Figure 2b). However, given the orientation of the beam splitter at 90° from that used with eyepiece optics, the final image of the exit pupil after the beam splitter is coincident with the eyepoint of the eye as required. In the case of a virtual pupil, however, as the FOV increases, the optics size remains almost invariant. Furthermore, in the case of projection optics where the final pupil is virtual, it is quite straightforward to design the optics with the pupil located at or close to the nodal points of the lens (i.e., mathematical first order constructs with unit angular magnification). In such a case, there will be little or no distortion. Correcting for distortion eliminates the need for real-time distortion correction with software or hardware. This property holds for HMPD design with relatively large FOVs. While optical correction is currently readily available for various hardware solutions, eliminating the need for distortion correction not only minimizes the cost of the system, but also eliminates processing. Therefore, there are no additional system delays. Furthermore, distortion-free optics avoids deterioration in image quality that can become more pronounced with increasing amounts of required correction. The fact that the microdisplay pixels remain square, regardless of the level of predistortion compen-sation of the rendered image, may cause visual discomfort if the pixels are resolvable, as commonly encountered in HMDs. Thus, distortion-free optics, even today with high speed processing, presents key advantages to more natural perception in, generally speaking, HMDs.

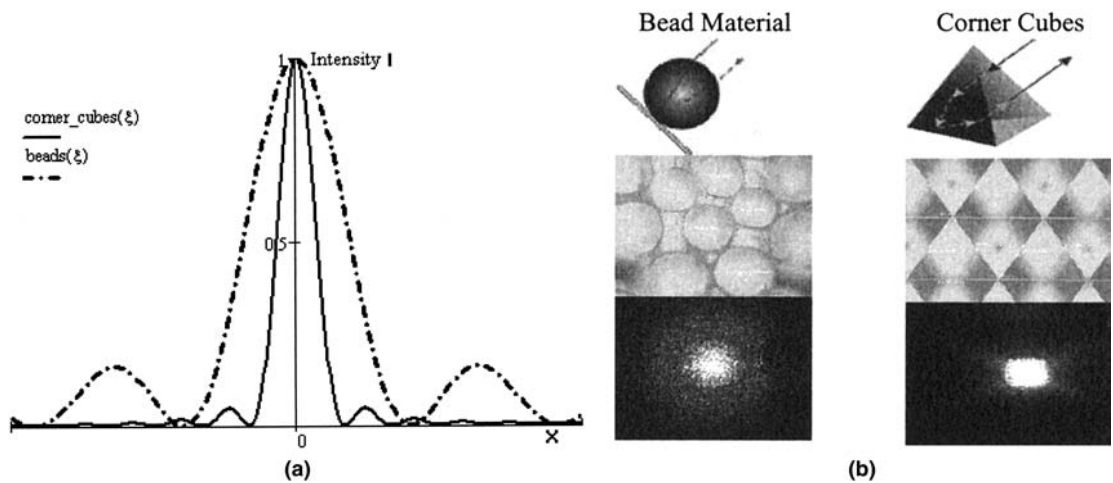

**Figure 4.** *(a) Theoretical modeling of the point spread function (PSF) for two kinds of retro-reflective materials (b) 3D (x,y,I) measured point spread functions; a microscopic view of the two different types of materials is shown above the PSFs, as well as their basic 3D microstructure.*

### 3.2 The Retro-Reflective Screen: Any Shape in Any Location

A key property of retro-reflective material is that rays hitting the surface at any angle are reflected back onto the same incident optical path. Thus theoretically, the perception of the image is independent of the shape and location of the screen made of such material. Such technology can thus be implemented with curved or tilted screens with no apparent distortion. In practice, depending on the specifics of the retro-reflective material some dependence on the shape may be observed outside a range of bending of the material. Also, image quality (i.e., the sharpness of the image) is in practice limited by the optical diffraction of the material (Martins & Rolland, 2003), whose impact can be highly significant for large discrepancies in the location of the material with respect to the optical image projected by the optics. Thus ideally, to minimize the effect of diffraction, the material must be placed in close proximity to the images focused by the projector. Furthermore, depending on the specific properties of the material, the amount of blurring, quantified as the width of the point spread function (PSF), may be more or less pronounced for the same discrepancy in the location of the projected images and the material as now quantified.

An analysis of diffraction was conducted on two types of material, a micro-beaded type of material (i.e., Scotchlite 3M Fabric Silver-beaded), and a micro corner-cube type of material (i.e., Scotchlite 3M Film Silver-cubed), both shown on a microscopic scale in Figure 4b.[1] The analysis shown in Figure 4a quantifies the spread of light after retro-reflection due to diffraction. Measurements made in the laboratory and also reported in Figure 4b indicate a good correlation in the results with the mathematical predictions. From this analysis, it is shown that the micro corner-cube material will be superior in maximizing the light throughput and minimizing loss in resolution due to diffraction. In a companion paper, an analysis of human performance in resolving small details with both types of materials is reported (Fidopiastis, Furhman, Meyer, & Rolland, 2005).

---

1. Scotchlite is a registered trademark of the 3M company.

Finally, it is important to note that the FOV of the HMPD is invariant with the location of the material in the environment, and the apparent size of a 3D object obtained from generating stereoscopic images will also be invariant with the location of the user with respect to the material as long as the head of the user is tracked in position and orientation as in any AR application.

### 3.3 Optical Design of HMPDs

The optical design of any HMD, including the HMPD, is highly driven by the choice of the microdisplays, specifically their size, resolution, and means of self-emitting light or requiring illumination optics. The smaller the microdisplays, the higher the required power of the optics to achieve a given FOV, and thus the higher the number of elements required.

The microdisplays and associated electronics first available for this project were 1.35 in. diagonal backlighting color AM-LCDs (active matrix liquid crystal displays) with $(640 \cdot 3) \cdot 480$ pixels and 42-$\mu$m pixel size. While higher resolution would be preferred, the availability in size and color of this microdisplay were determinant for the choice made. A 52° FOV optics per eye with a 35-mm focal length optics was designed in order to maintain a visual resolution of less than about 4 arc minutes. This choice resulted in a predicted 4.1 arc minutes per pixel in angular resolution, horizontally and vertically (Hua, Ha, & Rolland, 2003). In spite of the lower resolution imposed by the microdisplay, larger FOVs optics were explored. For example, a 70° FOV optics was investigated (Ha & Rolland, 2004), together with a discussion of the properties of such optics (Rolland, Biocca, et al., 2004). Both designs were based on an ultralightweight four-element compact projection optics using a combination of DOEs, plastic components, and aspheric surfaces. While plastic components are ideal to design an ultralight system, its combination with glass components and DOEs also enables higher image quality. The total weight of each lens assembly was only 6 g. The mechanical dimensions of the 52° and 70° FOVs optics were 20-mm in length by 18-mm in diameter and 15-mm in length by 13.4 mm in diameter, respectively. For both designs, an analysis of performance determined that the polychromatic modulation transfer functions displayed more than 40% contrast at 25 line-pairs/mm for a full size pupil of 12 mm. The distortion was constrained to be less than 2.5% across the overall visual fields in both cases.

Finally, whether the microdisplay is self-emitting or requires illumination optics may impose additional constraints on the design and compactness. If the microdisplay acts as a mirror, such as liquid crystal on silicon (LCOS) displays (Huang et al., 2002), the projection optics diameter will be larger than the microdisplay to avoid vignetting the footprint of the telecentric light beam reflected off the LCOS as it passes through the lens. Telecentric means that the central ray of the cone of light from each point in the field of view is parallel to the optical axis before the LCOS display. Furthermore, such microdisplays require illumination optics and thus they require additional optical and opto-mechanical components that will add weight, complexity, and cost to the overall system. Advantages of LCOS displays today, however, are their physical size (i.e., ~1 in. diagonal), their brightness, and their resolution.

For each HMPD design (and this also applies to HMD design in general), the choice of the microdisplay is critical to the success of a final prototype or product, and as such a choice must be application driven if the prototype or product are targeted at a real-world application.

## 4 Design of HMPD Technologies for Specific Applications: The Teleportal HMDP (T-HMPD) and the Mobile HMPD (M-HMPD)

Nonverbal cues regarding what others are thinking and where they are looking are key sources of information during real-time collaboration among workmates, especially when movements must be coordinated as in collaborative object manipulation. The direction of another's gaze is a key source of information as to the other's visual attention. For example, it helps disambiguate spatially ambiguous but common, everyday phrases such as "get the tool over there." Facial expressions supplemented by language provide teammates with insight to the intentions, moods, and meanings communicated by others.

Video conferencing systems provide some information on visual expressions, but fail to provide accurate cues of spatial attention and are poor at supporting physical collaboration because collaborators lack a common action space. VR systems using HMPD displays can better create a common action space and support naturalistic hand manipulation of 3D virtual objects during immersive collaboration, but facial expressions are partially occluded by the HMPD for local participants, and further masked for remote participants. A key challenge in immersive collaborative systems is how to add the important information channels of facial expressions and visual attention into a distributed AR interface. An emerging version of the HMPD design tailored for face-to-face interaction, and referred to as the teleportal HMPD (T-HMPD), will be described in Section 4.1.

Another challenge in creating distributed collaborative environments is how to create mobile systems based on HMPD technology, given that collaboration may also take place as we navigate through a real environment such as a museum, or a city. In such cases, it is not possible to position retro-reflective material strategically in the environment, however it would be advantageous based on the ultralightweight of the optics of HMPDs to expand the technology to mobile systems. A mobile HMPD (M-HMPD) will be described in Section 4.2.

### 4.1 The Teleportal HMPD (T-HMPD)

The T-HMPD integrates optical, image processing, and display technologies to capture, transport, and reconstruct an AR 3D model of the head and facial ex-pression of a remote collaborator networked to a local partner. The T-HMPD is a hybrid optical and video see-through HMD composed of an HMPD combined with a pair of lipstick video cameras and two miniature mirrors mounted to the side and slightly forward of the face of the user as shown schematically in Figure 5 (Biocca & Rolland, 2004). The configuration captures stereoscopic video of the user's face including both sides of the face without occlusion, with minimal interference within the user's visual field and only minimal occlusion of the face as viewed by other physical participants at the local site. Unlike room-based video, the head-worn camera and mirror system captures the full face of users no matter where they are looking and regardless of their locations, as long as the cameras are connected.

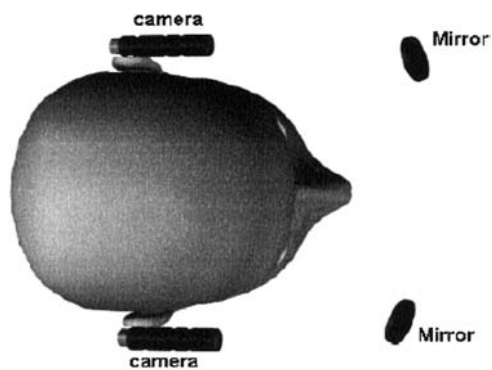

**Figure 5.** *General layout of the minimally intrusive stereoscopic face capture system.*

Figure 6a,b show the left and right views, respectively, of the lipstick cameras through the miniature mirrors (i.e., 1 in. in diameter in this case) in one of our first tests. The radius of curvature of the convex surface leading to the face capture shown was selected to be 65-mm from applying basic optical imaging equations between a small 4 mm focal length lipstick camera and the face. In the first implementation, the lipstick video cameras were Sony Electronics DXCLS1/1. Adjustable rods were designed to mount the two mirrors in order to experiment with various configurations of mirrors, camera lenses, and distances from the face and the two

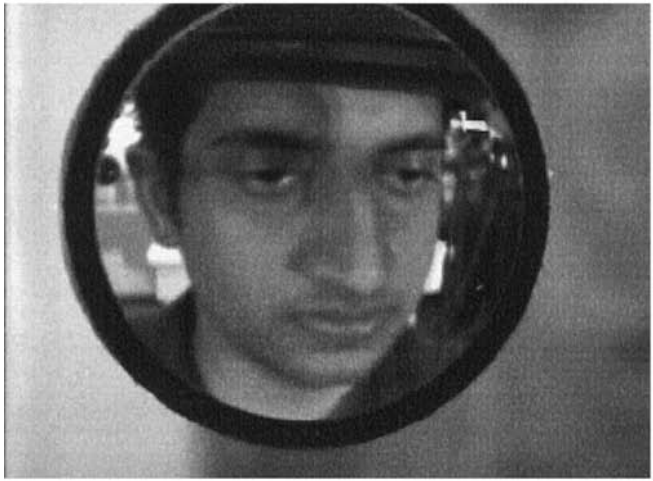

(a)

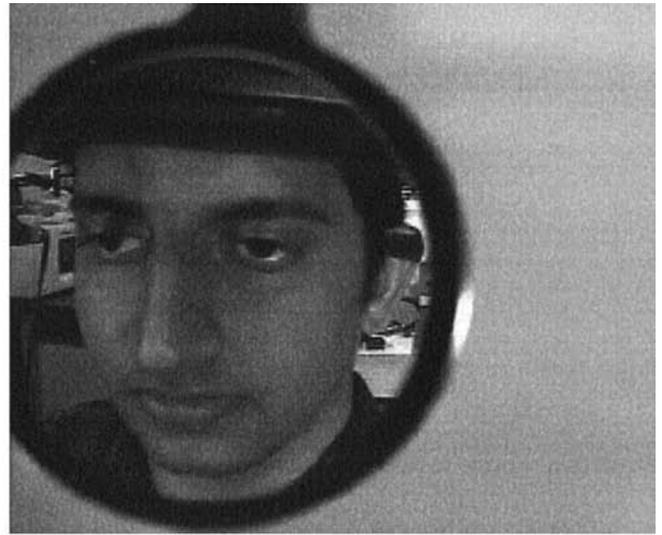

(b)

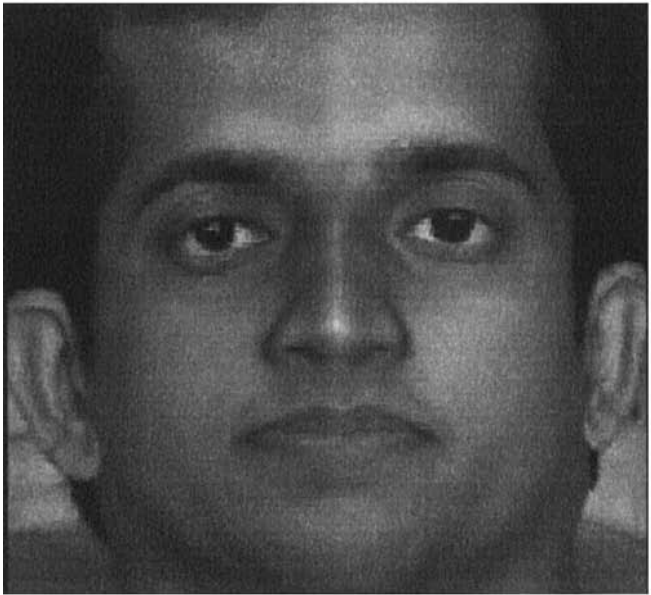

(c)

**Figure 6.** *Face images captured using the FCS (a) left image and (b) right image (c) Virtual frontal view generated from two side view cameras mounted in this case to a table as opposed to the head, for feasibility study (Courtesy frontal view from C. Reddy and G. Stockman, Michigan State University).*

mirrors. Optimization of the optical face capture system is under optimization for robustness, minimization of shadows on the face, and placement of the mirror side of the beam splitter to capture the face of any partici-pant with no interference from the projector light or the beam splitter. Image processing algorithms, under development at the MIND Lab (Media Interface and Network Design Lab) in collaboration with the ODA Lab (Optical Diagnostics and Applications Lab) unwrap the distorted images from the cameras and produce a composite video texture from the two images (Reddy, 2003; Reddy, Stockman, Rolland, & Biocca, 2004). A feasibility of creating a frontal virtual view from two lipstick cameras mounted to the side and slightly forward of the face is shown in Figure 6c, where the cameras in this case were mounted to a table. The composite stereo video texture can be sent directly via high bandwidth connection or mapped to a 3D mesh of the user's face. The 3D face of the user can be projected in the local space as a 3D AR object in the virtual environment and placed in the analogous spatial relation to others and virtual objects using the tracker data of the remote site. Alternatively the teleportal 3D face-to-face model can be projected onto a head model covered in retro-reflective material or as a 3D video on the wall of an office or conference room as in traditional teleconferencing systems. As the algorithm for stereo face capture and reconstruction matures, we are preparing to test the algorithm in various presentation scenarios including a retro-reflective ball and head model, and other embodiments of the remote others will be created. In optimizing the presentation of information to create the maximum sense of presence, we may investigate how to best display the remote faces. For example we may employ a retro-reflective tabletop where 3D scientific visualization may be shared, or a retro-reflective wall that would open a common window to both distributed visualization and social environments.

For local participants wearing HMPDs, a participant would see the eyes of the other participant illuminated; however the projector is too weak to shine light in the eyes of a side participant. The illuminated eyes may be turned off automatically when two local participants speak face-to-face to each other (while their face is still being captured), and the projector can be automatically turned back on when facing any retro-reflective material. This feature has not yet been implemented. There is no such issue for remote participants, given how the face is captured.

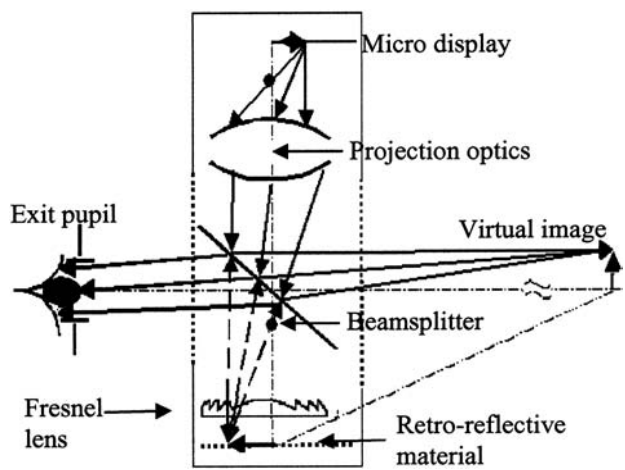

**Figure 7.** *Conceptual design of a mobile HMPD (M-HMPD).*

## 4.2 Mobile Head-Mounted Projection Display (M-HMPD)

In order to expand the use of HMPD to applications that may not be able to allow placing of retro-reflective material in the environment, a novel display based on the HMPD was recently conceived (Rolland, Martins, & Ha, 2003). A schematic of the display is shown in Figure 7. The main novelty of the display lies in removing the fabric from the environment and solely integrating the retro-reflective material within the HMPD for imaging using additional optics between the beam splitter and the material to optically image the material at a remote distance from the user. A preferred location for the material is in coincidence with the monocular virtual images of the HMPD to minimize the effects of diffraction imposed by the microstructures of the optical material. Without the additional optics, we estimated that in the best case scenario visual acuity would have been limited to about 10 minutes of arc even with a finer resolution of the microdisplay, which would be visually unacceptable. Thus when the material within the HMPD is not used simply for increased illumination, the imaging optics between the integrated material and the beam splitter is absolutely required in order to provide adequate overall resolution of the viewed images. The additional optics is illustrated in

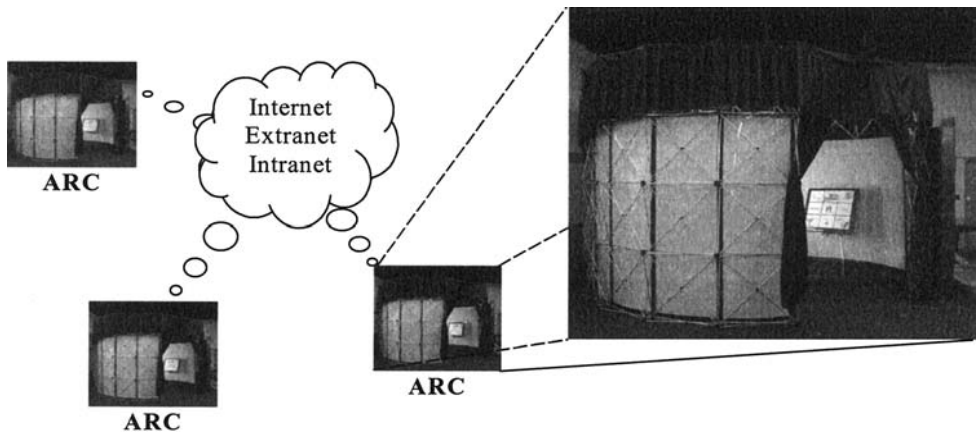

**Figure 8.** *Networked open environments with artificial reality centers (NOEs ARCs).*

Figure 7 with a Fresnel lens for low weight and compactness.

Because the image formed by the projection optics is minimized at the location of the retro-reflective material placed within the M-HMPD, a difference with the basic HMPD is that smaller microstructures are required (i.e., on the order of 10 $\mu$m instead of 100 $\mu$m). The detailed optical design of the lens is similar to that of the first ODA Lab HMPD optics except that the microdisplay chosen is a 0.6 in. diagonal OLED, and higher resolution 800 $\times$ 600 pixels. Details of the optical design were reported elsewhere (Martins, Ha, & Rolland, 2003). A challenge associated with the development of the M-HMPD is the design and fabrication of about 10 $\mu$m scale microstructure retro-reflective material. Such microstructures are not commercially available and custom-design materials are being investigated.

Because of its stand-alone capability, the M-HMPD extends the use of HMPDs to clinical guided surgery, medical simulation and training, wearable computers, mobile secure displays and distributed collaborative displays, as well as outdoor augmented see-through virtual environments. The visual results of the M-HMPD compare in essence to that of eyepiece-based see-through AR systems currently available. The difference lies however in the higher image quality (i.e., lower blur), a distortion-free image, and a more compact optics.

## 5 A Review of Collaborative Applications: Medical Applications and Infospaces

The HMPD facilitates the development of collaborative environments that allow seamless transitions through different levels of immersion from AR to a full virtual reality experience (Milgram & Kishino, 1994; Davis et al., 2003). In the artificial reality center (ARC), users can navigate between various levels of immersion that occur on the basis of where users position themselves with respect to the retro-reflective material. The ARC presented at ISMAR 2002 (Hamza-Lup, et al. 2002) together with a remote collaboration application built on top of DARE (distributed augmented reality environment) (Hamza-Lup, Davis, Hughes, & Rolland, 2002), consists of a generally curved or shaped, retro-reflective wall, an HMPD with miniature optics, a commercially available optical tracking system, and Linux-based PCs. The ARCs may take different shapes and sizes and are importantly quickly deployable either indoors or outdoors. Since the year 2001, we built two deployable (~10 minute setup) displays 10-ft wide by 7-ft high, as well as a deployable (~30 minute setup) 15-ft diameter center networked to some of the other ARCs as shown in Figure 8. However all displays aim to provide multiuser capability including remotely located users (Hamza-Lup & Rolland, 2004b), as well as 3D multisensory visualization including 3D sound, haptic, and smell (Rolland, Stanney, et al., 2004). The user's motion becomes the computer interface device, in contrast to systems that may resort to various display devices (Billinghurst, Kato, & Poupyrev, 2001).

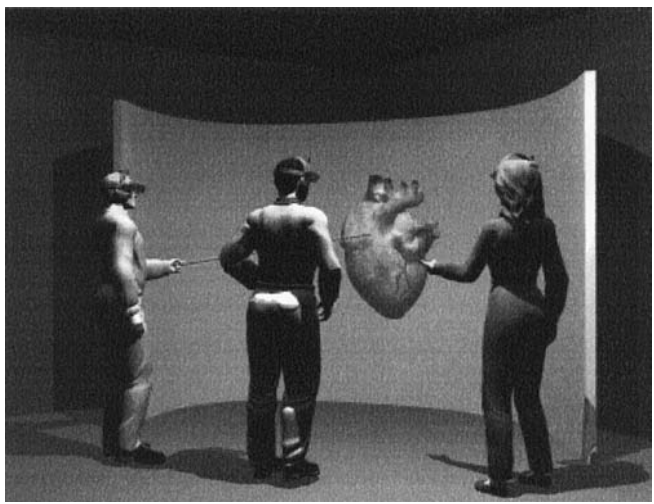

**Figure 9.** *Concept of multiusers interacting in the ARC.*

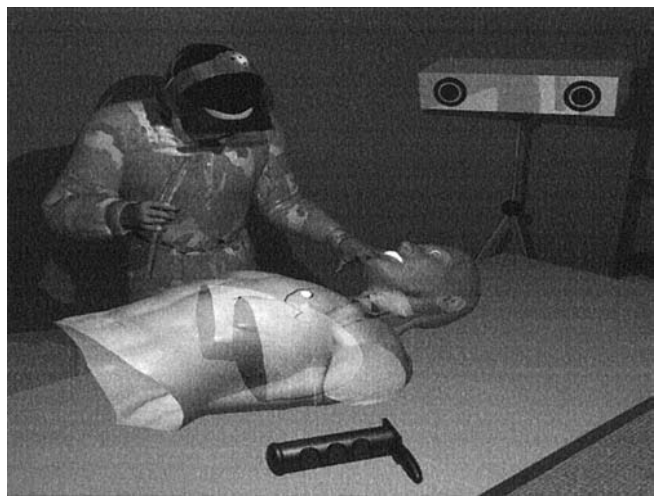

**Figure 10.** *Illustration of the endotracheal intubation training tool.*

Variants of the HMPD optics targeted at specific applications such as 3D medical visualization and distributed AR medical training tools (Rolland, Hamza-Lup, et al., 2003; Hamza-Lup, Rolland, & Hughes, 2005), embedded training display technology for the Army's future combat vehicles (Rodrigez, Foglia, & Rolland, 2003), 3D manufacturing design collaboration (Rolland, Biocca, et al., 2004) have been developed in our laboratory. Also, the HMPD in its original prototype form with a 52° ultralightweight optics, lies at the core of the Aztec Explorer application developed at the 3DVIS Lab that investigates various interaction schemes within SCAPE (stereoscopic collaboration in augmented and projective environments) that consists of an integration of the HMPD, together with the HiBall3000 head tracker by 3rd Tech (www.3rdtech.com), a SCAPE API developed in the 3DVIS Lab, the CAVERN G2 API networking library (www.openchanelsoftware.org), and a tracked 5DT Dataglove (Hua, Brown, & Gao, 2004).

In summary, the HMPD technology we have developed currently provides a fine balance of affordability and unique capabilities such as: (1) spanning the virtuality continuum allowing both full immersion and mixed reality, which may open a set of new capabilities across various applications, (2) enabling teleportal capability with face-to-face interaction, (3) creating ultralightweight wide FOVs mobile and secure displays, (4) creating small, low cost desktop technology or on larger scale quickly deployable 3D visualization centers or displays.

## 5.1 Medical Visualization and Augmented Reality Medical Training Tools

In the ARCs, multiple users may interact on the visualization of 3D medical models as shown in Figure 9, or practice procedures on AR human patient simulators (HPS) with a teaching module referred to as the ultimate intubation head (UIH) as shown in Figure 10, which we shall now further detail. Because as detailed earlier in the paper the light projected from one user returns only to that user, there is no cross talk between users, and thus multiple users can coexist with no overlapping graphics in the ARCs.

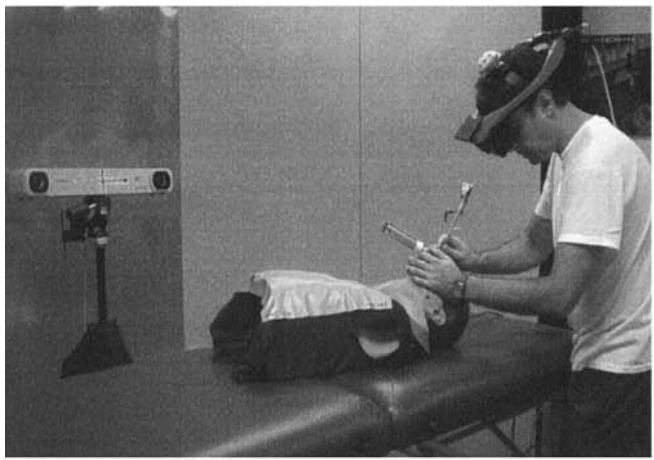
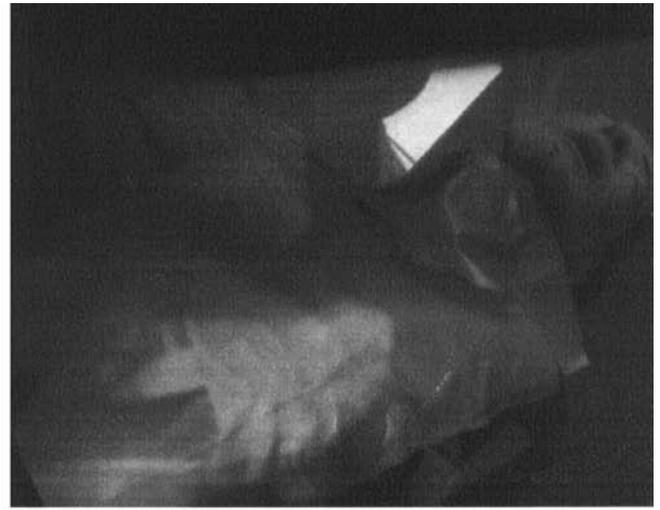

**(a)**                                    **(b)**

**Figure 11.** *(a) A user training on an intubation procedure (b) Lung model superimposed on the HPS using AR. (3D model of the Visible Human Dataset lung, Courtesy of Celina Imielinska, Columbia University.)*

The UIH is a training tool in development in the ODA Lab for endotracheal intubation (ETI) based on HMPD and ARC technology (Rolland, Hamza-Lup, et al., 2003) (Hamza-Lup, Rolland, & Hughes, 2005). It is aimed at medical students, residents, physician assistants, pre-hospital care personnel, nurse-anesthetists, experienced physicians, and any medical personnel who need to perform this common but critical procedure in a safe and rapid sequence.

The system trains a wide range of clinicians in safely securing the airway during cardiopulmonary resuscitation as ensuring immediate ventilation and/or oxygenation is critical for a number of reasons. Firstly, ETI, which consists of inserting an endotracheal tube through the mouth into the trachea and then sealing the trachea so that all air passes through the tube, is a lifesaving procedure. Secondly, the need for ETI can occur in many places, in and out of the hospital. Perhaps the most important reason for training clinicians in ETI, however, is the inherent difficulty associated with the procedure (American Heart Association, 1992; Walls, Barton, & McAfee, 1999).

Current teaching methods lack flexibility in more than one sense. The most widely used model is a plastic or latex mannequin commonly used to teach advanced cardiac life support (ACLS) techniques, including airway management. The neck and oropharynx are usually difficult to manipulate without inadvertently "breaking" the model's teeth or "dislocating" the cervical spine, because of the awkward hand motions required. A relatively recent development is the HPS, a mannequin-based simulator. The HPS is similar to the existing ACLS models, but the neck and airway are often more flexible and lifelike, and can be made to deform and relax to simulate real scenarios. The HPS can simulate heart and lung sounds and provide palpable pulses as well as realistic chest movement. The simulator is interactive, but requires real-time programming and feedback from an instructor (Murray & Schneider, 1997). Utilizing a HPS combined with 3D AR visualization of the airway anatomy and the endotracheal tube, paramedics will be able to obtain a visual and tactile sense of proper ETI. The UIH will allow paramedics to practice their skills and provide them with the visual feedback they could not obtain otherwise.

Intubation on the HPS is shown in Figure 11a. The location of the HPS, the trainee, and the endotracheal tube are tracked during the visualization. The AR system integrates an HMPD and an optical tracker with a Linux-based PC to visualize internal airway anatomy optically superimposed on a HPS as shown in Figure 11b. In this implementation, the HPS wears a custom-made T-shirt made of retro-reflective material. With the exception of the HMPD, the airway visualization is realized using commercially available hardware components. The computer used for stereoscopic rendering has a Pentium 4 CPU running a Linux-based OS with GeForce 4 GPU. The locations of the user, the HPS, and the endotracheal tube are tracked using a Polaris hybrid optical tracker from Northern Digital Inc. and custom designed probes (Davis, Hamza-Lup, & Rolland, 2004).[2]

In an effort to develop the UIH system, we had to acquire high quality textured models of anatomy (e.g., models from the Visible Human Dataset), develop methods for scaling these models to the HPS, as well as methods for registration of virtual models with real landmarks on the HPS. Furthermore, we are working towards interfacing the breathing HPS with a physiologically-based 3D real-time model of breathing lungs (Santhanam, Fidopiastis, Hamza-Lup, & Rolland, 2004). In the process of developing such methods, we are using the ARCs for the development and visualization of the models. Based on recent algorithms for dynamic shared state maintenance across distributed 3D environments (Hamza-Lup & Rolland, 2004b), we plan before the end of year 2005 to be able to share the development of these models in 3D and in real time with Columbia University Medical School, which has been working with us as part of a related project on decimating the models for real-time 3D visualization and will further be working with us on the testing of bringing these models alive (e.g., a deformable, breathing 3D model of the lungs).

2. Polaris is a registered endmark of Northern Digital Inc.

## 5.2 From Hardware to Interfaces: Mobile Infospaces Model for AR Menu, Tool, Object, and Data Layouts

The development of HMPDs and mobile AR systems allows users to walk around and interact with 3D objects, to be fully immersed in virtual environments, but also remain able to see and interact with other users located nearby. 3D objects and 2D information overlays such as labels can be tightly integrated with the physical space, the room, and the body of the user. In AR environments *space is the interface*. AR virtual spaces can make real world objects and environments rich with "knowledge" and guidance in the form of overlaid and easily accessible virtual information (e.g., labels of "see through" diagrams) and capabilities.

A research project called *mobile infospaces* at the M.I.N.D. Lab seeks general principles and guidelines for AR information systems. For example, how should AR menus and information overlays be laid out and organized in AR environments? The project focuses on what is novel about mobile AR systems: how should menus and data be organized around the body of a fully mobile user accessing high volumes of information?

Before optimizing AR we felt it was important to start by asking a fundamental question: Can AR environments improve individual and team performance in navigation, search, and object manipulation tasks when compared to other media? To answer this question we conducted an experiment to test whether spatially registered AR diagrams and instructions could improve human performance in an object assembly task. We compared spatial registered AR interfaces to three other media interfaces that presented the exact same 3D assembly information including computer aided instruction (standard screen), a printed manual, and non-spatially registered AR (i.e., 3D instructions on a see-through HMD). Compared to other interfaces, spatially registered AR was dramatically superior, reducing error rates by as much as 82%, reducing the user's sense of cognitive load between 5–25%, and speeding the time to complete the assembly task by 5–26% (Tang, Owen, Biocca, & Mou, 2003). These improvements in performance will vary with tasks and environments, but the controlled comparison suggests that the spatial registration of information provided by AR can significantly affect user performance.

If spatially registered AR systems can help improve human performance in some applications, then how can designers of AR interfaces optimize the placement and organization of virtual information overlays? Unlike the classic windows desktop interface, systematic principles and guidelines for the full range of menu, tool, and data object design aimed at mobile and team based AR systems are not well established. [See for example the useful but incomplete guidelines by Gabbard and Hix (2001).]

Our mobile infospaces use a model of human spatial cognition to derive and test principles and associated guidelines for organizing and displaying AR menus, object clusters, and tools. The model builds upon neuropsychological and behavioral research on how the brain keeps segments, organizes, and tracks the location of objects and agents around the body (egocentric space) and the environment (exocentric space) as a user interacts with the environment (Bryant, 1992; Cutting & Vishton, 1995; Grusser, 1983; Previc, 1998; Rizzolatti, Gentilucci, & Matelli, 1985). The mobile infospaces model seeks to map some key cognitive properties of the space around the body of the user to provide guidelines for the creation of AR menus and information tools.

Using AR can help users keep track of high volumes of virtual objects such as tools and data objects attached like an egocentric framework (i.e., field) around the body. People have a tremendous ability to easily keep track of objects in the environment, known as spatial updating. In some experiments, exploring whether this capability could be leveraged for AR menus and data, we explored whether new users could adapt their automatic ability to update the location of objects in a field around the body. Could they quickly learn and remember the location of a field of objects organized around their moving body, that is, virtual objects floating in space but affixed as tools to the central axis of the body? Even though such a task was new and somewhat unnat-ural, users were able to update the location of virtual objects attached to a framework around the body with as little as 30 seconds of experience or by simply being told that the objects would move (Mou, Biocca, Owen, et al., 2004). This finding suggests that users of AR systems might be able to quickly keep track of and access many tools and objects floating in space around their body, even though fields of floating objects attached to the body is a novel model and not experienced in the natural world because of the simple laws of gravity.

If users can make use of tools and menus freely moving around the body, then are there sweet spot locations around the body that may have different cognitive properties? Some areas are clearly more attention getting, but they may also have slightly different ergonomics, different memory properties, or even slightly different meaning (semantic properties). Basic psychological research indicates that locations in physical and AR space are by no means psychologically equal; the psychological properties of space around the body and the environment are highly asymmetrical (Mou, Biocca, Tang, & Owen, 2005). Perception, attention, meaning, and memory for objects can vary with their locations and organization in egocentric and exocentric space.

In a program to map some of the static and dynamic properties of the spatial location of virtual objects around a moving body, we conducted a study of the ergonomics of the layout of objects and menus. The experiment found that the speed for which a user wearing an HMD can find and place a virtual tool in the space around the body can vary by as much as 300% (Biocca, Eastin, & Daugherty, 2001). This finding has implications on where to place frequently used tools and objects. Locations in space, especially those around the body, can also have different psychological properties. For example, a virtual object, especially agents (i.e., virtual faces), may be perceived with slightly different shades of meaning (i.e., connotations) as their location around the body varies (Biocca, David, Tang, & Lim, 2004). Because the meaning of faces varied strongly, there are implications for where in the space around the body designers might place video-conference windows or artificial agents.

Current research on the mobile infospaces project is expanding to produce: (1) a map of psychologically relevant spatial frames, (2) a set of guidelines based on existing research and practices, and (3) an experiment to explore how spatial organization of information can augment or support human cognition.

## 6    Summary of Current Properties of HMPDs and Issues Related to Future Work

*Integrating immersive physical and virtual spaces.* Many of the fundamental issues in the design of collaborative environments deal with the representation and use of space. HMPDs can combine the wide-screen immersive experience of CAVE with the hands on, close to the body immediacy of see-through, head-worn AR systems. The ARCs constitute an example of the use of the technology in wide-screen immersive environments, which are quickly deployable indoors or outdoors. Various applications in collaborative visualization may require simultaneous display of large immersive spaces such as rooms and landscapes with more detailed handheld spaces such as models and tools.

With HMPDs, each person has a unique perspective on both physical and virtual space. Because the information comes from each user's display, information seen by each user such as labels can be unique to that user, allowing for customization and private information within a shared environment. The future development of HMPDs shares some common issues with other HMDs such as resolution, luminosity, FOV, comfort, and also with other AR systems such as registration accuracy. Beyond these, issues such as those addressed by our mobile infospaces program seek to model the space afforded by HMPDs as an integrated information environment making full use of the HMPDs ability to integrate information and the faces of collaborators around the body and the environment.

*AR and collaborative support.* Like CAVE technology or other display systems, the HMPD approach supports users experiencing among other wide-screen, immersive

3D visualizations and environments. But unlike CAVE, the projection display is head-worn, providing each user with a correct perspectives viewpoint on the virtual scene, essential to the display of objects that are to be hand manipulated close to the body. Continued development of these features needs to consider, as with other AR systems, continued registration of virtual and physical objects especially in the context of multiple users. The mobile infospaces research program explores guidelines for integrating and organizing physical and virtual tools.

*Immersive-AR properties.* The properties of the HMPD with retro-reflective surfaces support AR properties of any object that incorporates some retro-reflective material. For example, this property can be used in immersive applications to allow projection walk-around displays such as the visualization of a full body in a display tube-screen or on handheld objects such as tablets. Unlike see-through optical displays, the AR objects produced via a HMPD have some of the properties of visual occlusion. For example, physical objects appearing in front of the virtual object will occlude it as shown in Figure 3. Because the virtual images are attached to physical objects with retro-reflective surfaces, users view the space immediately around their body, but they can also move around, pick up, and interact with physical objects on which virtual information such as labeling, color, and other virtual properties can be annotated.

*Designing spaces for multiple collaborative users.* Collaborative spaces need to support spatial interaction among active users. The design of T-HMPDs seeks to minimize obscuring the face of the user to other local users in the physical space, a problem in VR HMDs, while still immersing the visual system in a unique, perspective accurate, immersive AR experience. The T-HMPD attempts to integrate some of the social cues from two distributed collaborative spaces into one common space. Much research is underway to develop the T-HMPD to achieve real-time distributed face-to-face interactions, a promise of the technology, and testing its benefit compared to 2D video streaming.

*Mobile spaces.* While the HMPD originally was pro-

posed to capitalize on retro-reflective material strategically positioned in the environment, we have extended the display concept to provide a fully mobile display, the M-HMPD. This display still uses projection optics, but together with retro-reflective material fully embedded into the HMPD, opens a breadth of new applications in natural environments. Future work includes fabricating the custom-designed micro retroreflective material needed for the M-HMPD and expanding on the FOV for various applications.

# 7. Conclusion

In this paper we have reviewed a novel emerging technology, the HMPD, based on custom designed miniature projectors mounted on the head and coupled with retro-reflective material. We have demonstrated that such material may be positioned either in the environment at strategic locations or within the HMPD itself, leading to full mobility with M-HMPDs. With the development of HMPDs, spaces such as the quickly deployable augmented reality centers (ARCs) and mobile AR systems allow users to interact with 3D objects and other agents located around a user or remotely. Yet another evolution of the HMPD, the teleportal T-HMPD, seeks to minimize obscuring the face of the user to other local users in the physical space in order to support spatial interaction among active users, while also providing remote users a potential face-to-face collaboration with remote participants. Projection technologies create virtual spaces within physical spaces. In some ways we can see the approach to the design of HMPDs and related technologies as a program to integrate various issues in representation of AR spaces and integrate distributed spaces into one fully embodied, shared collaborative space.

# Acknowledgments

We thank NVIS Inc. for providing the opto-mechanical design of the HMPD shown in Figure 1c and Figure 10. We thank 3M Corporation for the generous donation of the cubed retro-reflective material. We also thank Celina Imielinska from Columbia University for providing the models from the Visible Human Dataset displayed in this paper. We thank METI Inc. for providing the HPS and stimulating discussions about medical training tools. This research started with seed support from the French ELF Production Corporation and the M.I.N.D. Lab at Michigan State University, followed by grants from the National Science Foundation IIS 00-82016 ITR, EIA-99-86051, IIS 03-07189 HCI, IIS 02-22831 HCI, the US Army STRICOM, the Office of Naval Research contracts N00014-02-1-0261, N00014-02-1-0927, and N00014-03-1-0677, the Florida Photonics Center of Excellence, METI Corporation, Adastra Labs LLC, and the LINK and MSU Foundations.

# References

American Heart Association. (1992). Guidelines for cardiopulmonary resuscitation and emergency cardiac care—Part II: Adult Basic Life Support. *Journal of the American Medical Association, 268,* 2184–2198.

Billinghurst, M., Kato, H., & Poupyrev, I. (2001). The magicbook: Moving seamlessly between reality and virtuality. *IEEE Computer Graphics and Applications (CGA), 21*(3), 6–8.

Biocca, F., & Rolland, J.P. (2004, August). *US Patent No. 6,774,869.* Washington, DC: U.S. Patent and Trademark Office.

Biocca, F., David, P., Tang, A., & Lim, L. (2004, May). *Does virtual space come precoded with meaning? Location around the body in virtual space affects the meaning of objects and agents.* Paper presented at the 54th Annual Conference of the International Communication Association, New Orleans, LA.

Biocca, F., Eastin, M., & Daugherty, T. (2001). *Manipulating objects in the virtual space around the body: Relationship between spatial location, ease of manipulation, spatial recall, and spatial ability.* Paper presented at the International Communication Association, Washington, DC.

Biocca, F., Harms, C., & Burgoon, J. (2003). Towards a more robust theory and measure of social presence: Review and suggested criteria. *Presence: Teleoperators and Virtual Environments, 12*(5), 456–480.

Bryant, D. J. (1992). A spatial representation system in humans. *Psycholoquy, 3*(16), 1.

Cruz-Neira, C. D., Sandin, J., & DeFanti, T. A. (1993, August). Surround-screen projection-based virtual reality: The design and implementation of the CAVE. *Proceedings of ACM SIGGRAPH 1993,* (pp. 135–142). Anaheim, CA.

Cutting, J. E., & Vishton, P. M. (1995). Perceiving layout and knowing distances: The integration, relative potency, and contextual use of different information about depth. In W. Epstein & S. Rogers (Eds.), *Perception of space and motion* (pp. 69–117). San Diego, CA: Academic Press.

Davis, L., Hamza-Lup, F., & Rolland, J. P. (2004). A method for designing marker-based tracking probes. *International Symposium on Mixed and Augmented Reality ISMAR '04* (pp. 120–129), Washington, DC.

Davis, L., Rolland, J. P., Hamza-Lup, F., Ha, Y., Norfleet, J., Pettitt, B., et al. (2003). Enabling a continuum of virtual environment experiences. *IEEE Computer Graphics and Applications (CGA), 23*(2), 10–12.

Fidopiastis, C. M., Furhman, C., Meyer, C., & Rolland, J. P. (2005). Methodology for the iterative evaluation of prototype head-mounted displays in virtual environments: Visual acuity metrics. *Presence: Teleoperators and Virtual Environments, 14*(5) 550–562.

Fergason, J. (1997). *US Patent No. 5,621,572.* Washington, DC: U.S. Patent and Trademark Office.

Fisher, R. (1996). *US Patent No. 5,572,229.* Washington, DC: U.S. Patent and Trademark Office.

Gabbard, J. L., & Hix, D. (2001). Researching usability design and evaluation guidelines for Augmented Reality Systems. Retrieved May 1, 2004 from www.sv.vt.edu/classes/ESM4714/Student_Proj/class00/gabbard/.

Girardot, A., & Rolland, J. P. (1999). *Assembly and investigation of a projective head-mounted display* (Technical Report TR-99-003). Orlando, FL: University of Central Florida.

Grusser, O. J. (1983). Multimodal structure of the extrapersonal space. In A. Hein & M. Jeannerod (Eds.), *Spatially oriented behavior* (pp. 327–352). New York: Springer.

Ha, Y., & Rolland, J. P. (2004, October). *US Patent No. 6,804,066 B1.* Washington, DC: U.S. Patent and Trademark Office.

Hamza-Lup, F., Davis, L., Hughes, C., & Rolland, J. (2002). Where digital meets physical—Distributed augmented reality environments. *ACM Crossroads 2002,* 9(3).

Hamza-Lup, F. G., Davis, L., & Rolland, J. P. (2002, September). The ARC display: An augmented reality visualiza-

tion center. *International Symposium on Mixed and Augmented Reality (ISMAR),* Darmstadt, Germany. Retrieved May 17, 2005 from http://studierstube.org/ismar2002/demos/ismar_rolland.pdf.

Hamza-Lup, F. G., & Rolland, J. P. (2004a, March). Adaptive scene synchronization for virtual and mixed reality environments. *Proceedings of IEEE Virtual Reality 2004* (pp. 99–106). Chicago, IL.

Hamza-Lup, F. G., & Rolland, J. P. (2004b). Scene synchronization for real-time interaction in distributed mixed reality and virtual reality environments. *Presence: Teleoperators and Virtual Environments, 13*(3), 315–327.

Hamza-Lup, F. G., Rolland, J. P., & Hughes, C. E. (2005). A distributed augmented reality system for medical training and simulation. In B. J. Brian (Ed.), *Energy, simulation-training, ocean engineering and instrumentation: Research papers of the link foundation fellows,* (Vol. 4, pp. 213–235). Rochester, NY: Rochester Press.

Hua, H., Brown L. D., & Gao, C. (2004). System and interface framework for SCAPE as a collaborative infrastructure. *Presence: Teleoperators and Virtual Environments 13*(2), 234–250.

Hua, H., Girardot, A., Gao, C., & Rolland, J. P. (2000). Engineering of head-mounted projective displays. *Applied Optics, 39*(22), 3814–3824.

Hua, H., Ha, Y., & Rolland, J. P. (2003). Design of an ultra-light and compact projection lens. *Applied Optics, 42*(1), 97–107.

Hua, H., & Rolland, J. P. (2004). *US Patent No. 6,731,434 B1.* Washington, DC: U.S. Patent and Trademark Office.

Huang, Y., Ko, F., Shieh, H., Chen, J., & Wu, S. T. (2002). Multidirectional asymmetrical microlens array light control films for high performance reflective liquid crystal displays. *SID Digest,* 869–873.

Inami, M., Kawakami, N., Sekiguchi, D., Yanagida, Y., Maeda, T., & Tachi, S. (2000). Visuo-haptic display using head-mounted projector. *Proceedings of IEEE Virtual Reality 2000* (pp. 233–240), Los Alamitos, CA.

Kawakami, N., Inami, M., Sekiguchi, D., Yangagida, Y., Maeda, T., & Tachi, S. (1999). Object-oriented displays: A new type of display systems—From immersive display to object-oriented displays. *IEEE International Conference on Systems, Man, and Cybernetics* (Vol. 5, pp. 1066–1069). Piscataway, NJ.

Kijima R., & Ojika, T. (1997). Transition between virtual environment and workstation environment with projective head-mounted display. *Proceedings of IEEE 1997 Virtual Reality Annual International Symposium* (pp. 130–137), Los Alamitos, CA.

Krueger, M. (1977). Responsive environments. *Proceedings of the National Computer Conference* (pp. 423–433). Montvale, NJ.

Krueger, M. (1985). VIDEOPLACE—An artificial reality. *Proceedings of ACM SIGCHI'85* (pp. 34–40). San Francisco, CA.

Martins, R., Ha, Y., & Rolland, J. P. (2003). Ultra-compact lens assembly for a head-mounted projector. UCF Patent filed, University of Central Florida.

Martins, R., & Rolland, J. P. (2003). Diffraction properties of phase conjugate material. In C. E. Rash, and E. R. Colin (Eds.) *Proceedings of the SPIE Aerosense: Helmet- and Head-Mounted Displays VIII: Technologies and Applications. 5079,* 277–283.

Milgram P., & Kishino, F. (1994). A taxonomy of mixed reality visual displays. *IECE Trans. Information and Systems (Special Issue on Networked Reality), E77-D*(12), 1321–1329.

Mou, W., Biocca, F., Owen, C. B., Tang, A., Xiao, F., & Lim, L. (2004). Frames of reference in mobile augmented reality displays. *Journal of Experimental Psychology: Applied, 10*(4), 238–244.

Mou, W., Biocca, F., Tang, A., & Owen, C. B. (2005). Spatialized interface design of mobile augmented reality systems. *International Journal of Human Computer Studies.*

Murray W. B., & Schneider, A. J. L. (1997). Using simulators for education and training in anesthesiology. *ASA Newsletter, 6*(10). Retrieved May 1, 2003 from http://www.asahq.org/Newsletters/1997/10_97/Simulators_1097.html.

Parsons, J., & Rolland, J. P. (1998). A non-intrusive display technique for providing real-time data within a surgeon's critical area of interest. *Proceedings of Medicine Meets Virtual Reality 98* (pp. 246–251). Newport Beach, CA.

Poizat, D. & Rolland, J. P. (1998). *Use of retro-reflective sheets in optical system design* (Technical Report TR98-006). Orlando, FL: University of Central Florida.

Previc, F. H. (1998). The neuropsychology of 3D space. *Psychological Bulletin, 124*(2), 123–164.

Reddy, C. (2003). A non-obtrusive head mounted face capture system. Unpublished Master's Thesis, Michigan State University.

Reddy, C., Stockman, G. C., Rolland, J. P., & Biocca, F. A. (2004). Mobile face capture for virtual face videos. In *CVPR'04, The First IEEE Workshop on Face Processing in Video*. Retrieved July 7, 2004 from http://www.visioninterface.net/fpiv04/papers.html.

Rizzolatti, G., Gentilucci, M., & Matelli, M. (1985). Selective spatial attention: One center, one circuit, or many circuits? In M. I. Posner & O. S. M. Marin (Eds.), *Attention and performance* (pp. 251–265). Hillsdale, NJ: Erlbaum.

Rodriguez, A., Foglia, M., & Rolland, J. P. (2003). Embedded training display technology for the army's future combat vehicles. *Proceedings of the 24th Army Science Conference* (pp. 1–6), Orlando, FL.

Rolland, J. P., Biocca, F., Hua, H., Ha, Y., Gao, C., & Harrysson, O. (2004). Teleportal augmented reality system: Integrating virtual objects, remote collaborators, and physical reality for distributed networked manufacturing. In K. Ong & A.Y.C. Nee (Eds.), *Virtual and Augmented Reality Applications in Manufacturing* (Ch. 11, pp. 179–200). London: Springer-Verlag.

Rolland, J. P., & Fuchs, H. (2001). Optical versus video see-through head-mounted displays. In W. Barfield & T. Caudell (Eds.), *Fundamentals of Wearable Computers and Augmented Reality* (pp. 113–156). Mahwah, NJ: Erlbaum.

Rolland, J. P., Ha, Y., & Fidopiastis, C. (2004). Albertian errors in head-mounted displays: Choice of eyepoints location for a near or far field tasks visualization, *JOSA A, 21*(6), 901–912

Rolland, J. P., Hamza-Lup, F., Davis, L., Daly, J., Ha, Y., Martin, G., et al. (2003, January). Development of a training tool for endotracheal intubation: Distributed augmented reality. *Proceedings of Medicine Meets Virtual Reality* (Vol. 11, pp. 288–294). Newport Beach, California.

Rolland, J. P., Martins, R., & Ha, Y. (2003). Head-mounted display by integration of phase-conjugate material. UCF Patent Filed, University of Central Florida.

Rolland, J. P., Parsons, J., Poizat, D., & Hancock, D. (1998, July). Conformal optics for 3D visualization. Proceedings of the International Optical Design Conference 1998 SPIE, 3482, (pp. 760–764). Kona, HI.

Rolland, J. P., Stanney, K., Goldiez, B., Daly, J., Martin, G., Moshell, M., et al. (2004, July). Overview of research in augmented and virtual environments: RAVES. *Proceedings of the International Conference on Cybernetics and Information Technologies, Systems and Applications (CITSA'04), 4,* 19–24.

Santhanam, A., Fidopiastis, C, Hamza-Lup, F., & Rolland, J. P. (2004). Physically-based deformation of high-resolution 3D models for augmented reality based medical visualization.

*Proceedings of MICCAI-AMI-ARCS International Work-shop on Augmented Environments for Medical Imaging and Computer-Aided Surgery 2004* (pp. 21–32), Rennes/St. Malo, France.

Short, J., Williams, E., and Christie, B. (1976). Visual communication and social interaction. In R. M. Baecker (Ed.), *Readings in groupware and computer-supported cooperative work, assisting human-human collaboration* (pp. 153–164). San Mateo, CA: Morgan Kaufmann.

Sutherland, I. E. (1968). A head mounted three dimensional display. *Proceedings of the Fall Joint Computer Conference, AFIPS Conference, 33,* 757–764.

Tang, A., Owen, C., Biocca, F., & Mou, W. (2003). Comparative effectiveness of augmented reality in object assembly. *Proceedings of the Conference on Human Factors in Computing Systems ACM CHI* (pp. 73–80). Ft. Lauderdale, FL.

Walls, R. M., Barton, E. D., & McAfee, A. T. (1999). 2,392 emergency department intubations: First report of the ongoing national emergency airway registry study. *Annals of Emergency Medicine, 26,* 364–403.